%% file: climate-positive-blockchain-wp.tex
\documentclass[10pt,draftcls,onecolumn]{IEEEtran}
\usepackage{amsmath}
\usepackage{amssymb}
\usepackage{subfigure}
\usepackage{xcolor}
\usepackage{graphicx}

\ifCLASSINFOpdf
\else
\fi
\begin{document}
%
\title{Incentivizing Gigaton-Scale Carbon Dioxide Removal via a Climate-Positive Blockchain}
%
%
%

\author{Jonathan Bachman, Sujit Chakravorti, Shantanu Rane
        and~Krishnan Thyagarajan
\thanks{S. Rane, J. Bachman, and K. Thyagarajan are with Palo Alto Research Center (PARC, Inc.). S. Chakravorti is with Alliance for Innovative Regulation (AIR).}

}

\maketitle

\begin{abstract}
\noindent A new crypto token is proposed as an incentive mechanism to remove $\text{CO}_2$ from the atmosphere permanently at gigaton scale. The token facilitates $\text{CO}_2$ removal (CDR) by providing financial incentives to those that are removing $\text{CO}_2$ and an opportunity to provide additional financial resources for CDR by the public. The new token will be native to a blockchain that uses a Proof-of-Useful-Work (PoUW) consensus mechanism. The useful work will be conducted by direct air carbon capture and storage (DACCS) facilities that will compete with each other based on the amount of $\text{CO}_2$ captured and permanently stored. In terms of energy consumption, we require that the entire process, comprising DACCS technology and all blockchain operations, be climate positive while accounting for life cycle analysis of equipment used. We describe the underlying reward mechanism coupled with a verification mechanism for CDR. In addition, we consider security features to limit attacks and fraudulent activity. Finally, we outline a roadmap of features that are necessary to fully implement and deploy such a system, but are beyond the current scope of this article.  
\end{abstract}


%
\IEEEpeerreviewmaketitle

\input{intro-and-bkgnd}
\input{direct-air-capture}
\input{incentive-mech}

\input{consensus-mech}
\input{roadmap}



\section*{Acknowledgments}
We gratefully acknowledge the support received from the Alliance for Innovative Regulation (AIR) and Palo Alto Research Center (PARC). Insightful discussions with fellow participants at a January 2023 Tech Sprint organized by AIR and hosted by AlixPartners at their San Francisco Bay Area offices are also appreciated. In addition, we thank Dan Boneh and Paula Palermo for critical feedback on previous drafts. 

\ifCLASSOPTIONcaptionsoff
  \newpage
\fi

\bibliographystyle{unsrt}
\bibliography{references}

\end{document}

%% file: intro-and-bkgnd.tex
\section{Introduction and Background}
\label{sec:introduction}

As realization grows about the impact of climate change, policymakers have considered incentive mechanisms to reduce the rate of increase of carbon dioxide, ${\text{CO}}_2$, in the atmosphere. Such incentives include tax credits on the purchase of electric vehicles and subsidies for renewable energy generation. To reduce net ${\text{CO}}_2$ emissions, one incentive mechanism enables polluting agents to purchase carbon credits (also called offset credits) to compensate for their emissions of ${\text{CO}}_2$ and other greenhouse gases. While these credits can fund technologies to reduce the increase of ${\text{CO}}_2$ in the atomosphere, they have significant limitations. We are particularly concerned that carbon credits represent an easy option for polluters, removing the incentive to make the concerted changes necessary to mitigate climate change. 

In this article, we propose a mechanism that uses a crypto token to reward carbon dioxide removal (CDR) from the atmosphere using Direct Air Carbon Capture and Storage (DACCS).\footnote{CDR technologies do not include Carbon Capture and Storage (CCS) and Carbon Capture and Utilisation (CCU). According to Smith \textit{et al.} \cite{smith23stateofcdr}, to count as Carbon Dioxide Removal (CDR) technology, ``a method must be an intervention which captures $\text{CO}_2$ from the atmosphere and durably stores it.'' CCS and CCU are industrial processes aimed at reducing the emission of $\text{CO}_2$.} Unlike a carbon credit, that represents one metric ton of greenhouse gases removed from the atmosphere, our token does not represent an amount of $\text{CO}_2$ equivalent captured and stored. Our tokens are ``mined'' by the winning DACCS facility and its value is determined by the market for the token. Furthermore, the DACCS process including life-cycle analysis of the equipment used must be climate positive, namely capture more $\text{CO}_2$ equivalents than greenhouse gasses emitted. The adoption and usage of the token depends on the demand of two distinct groups: those that are willing to provide financial incentives for CDR, including individuals, businesses, and governments, and those that remove $\text{CO}_2$ from the atmosphere. Although not exact, one simple way to gauge the demand for CDR and the willingness to pay is to observe the demand for carbon credits. McKinsey estimates that the annual global demand for carbon credits could reach up to 1.5 to 2.0 gigatons of $\text{CO}_2$ by 2030 and the market size could range from \$5 billion to more than \$50 billion depending on various factors \cite{blaufelder2021blueprint}.  

To meet the climate goals of the 2015 Paris Agreement, we must substantially reduce greenhouse gas emissions, generate more energy from renewable sources, and capture and store $\text{CO}_2$ from the atmosphere at gigaton scale. Our proposed solution focuses on the last component. CDR approaches, such as reforestation, biochar, bioenergy with carbon capture and storage (BECCS) and DACCS, should be a part of the solution to limit global warming between 1.5°C or 2°C. Smith \textit{et al.} \cite{smith23stateofcdr} estimates that novel CDR approaches would have to increase by a factor ranging from 30 to 540 by 2030 to limit warming between 1.5°C or 2°C. 

For the DACCS CDR, we propose a solution that: verifies and records the quantity of $\text{CO}_2$ captured and stored by the winning DACCS facility on an immutable blockchain in real time, rewards DACCS facilities to remove and store gigatons of $\text{CO}_2$, and offers sufficient security against malicious attacks and fraud. Our results can be generalized to other CDR technologies as long as measurement of capture and storage can be done accurately and timely. 

We propose collecting telemetric data using tamper-resistant sensors and recording this information on an immutable climate-positive blockchain. In our case, we define a climate positive blockchain as the sum of $\text{CO}_2$ captured and stored, the amount of $\text{CO}_2$ equivalents emitted during the operation of the DACCS facility including life-cycle analysis of the equipment used, the $\text{CO}_2$ equivalents emitted from blockchain operations and storage of transactions along with life-cycle analysis of the equipment used.  The linking of the physical world with the digital world has its challenges including security especially guarding against adversarial attacks and fraudulent transactions. We propose a process that is highly automated and secure to provide sufficient reliability and trust in the data collected in real time. By doing so, we address a major challenge to the current system of carbon offsets, namely the transparency and verifiability of the amount of CDR and stored \cite{clisson23bnp}.

Our proposed blockchain uses Proof-of-Useful-Work (PoUW) consensus mechanism where the useful work is CDR.\footnote{For general discussion of PoUW consensus mechanisms, see \cite{baldominos2019coin}, \cite{hoffmann22arxiv}, and \cite{todorovic22symmetry}.} The winning DACCS facility would be determined by a weighted lottery draw based on the amount of $\text{CO}_2$ captured and stored. Because we require that each DACCS facility is on net removing $\text{CO}_2$ from the atmosphere, even the non-winning DACCS facility are contributing to the removal of $\text{CO}_2$. Upon verification by independent validators of the telemetric data, the winning DACCS facility would be financially rewarded with a quantity of tokens and would add the new block to the chain similar to a crypto miner participating in a PoW consensus mechanism. 

A market would develop where those that wanted to financially contribute to the removal of $\text{CO}_2$ could purchase these tokens from the winning facilities. These tokens represent verified $\text{CO}_2$ removal and storage. Clearly, the DACCS facilities would only expend additional resources to participate if they are able to convert these tokens into a medium of exchange, such as fiat currency, that could be used to buy goods and services. We also discuss how these tokens themselves could become a medium of exchange used to buy goods and services directly. In other words, these tokens would not need to be converted to fiat currency or other media of exchange.   

When compared with the enormous amount of energy consumed by the Bitcoin network to operate its Proof-of-Work-based consensus mechanism, our approach consumes a significantly smaller amount of energy. More importantly, while the large energy consumption in Bitcoin results in massive amounts of $\text{CO}_2$ emissions, our approach targets a massive net \emph{reduction} in the amount of atmospheric $\text{CO}_2$.\footnote{For estimates of the carbon footprint of Bitcoin, the most popular cryptocurrency using a PoW consensus mechanism, see \cite{deroche22sierra}, \cite{devries22joule}, and \cite{southpole22}. For example, according to Digiconomist website on July 7, 2023, the Bitcoin network consumes 102.45 terawatt-hours per year which is similar to annual power consumption of Kazakhstan and has a carbon footprint of 57.14 million tons of $\text{CO}_2$ per year which is similar to the carbon footprint of Portugal.} 

In other words, PoUW consensus mechanism that we propose has the advantage that the energy used to ``mine'' or add blocks to the blockchain is being used to remove net $\text{CO}_2$ from the atmosphere accounting for $\text{CO}_2$ equivalents used in the process. Crucially, \emph{all the miners} -- the DACCS facility that creates the next block, as well as all other DACCS facilities that do not win the competition to create the next block -- contribute to CDR. This should be contrasted with other PoW cryptocurrencies where the energy consumed by the unsuccessful miners is wasted.

This article is organized as follows. In Section~\ref{sec:direct-air-capture}, we describe the DACCS technology, why this technology is suitable for a PoUW consensus mechanisms, and identify the telemetry and mass balance equations that are necessary for the PoUW consensus mechanism. In Section~\ref{sec:incentive}, we outline the incentive mechanism using a decentralized framework. In Section~\ref{sec:consensus}, we describe the key steps in a notional protocol for establishing consensus in a climate positive blockchain and discuss some security considerations. In Section~\ref{sec:leveraging}, we explore how the token becomes a medium of exchange and used to purchase goods and services. In Section~\ref{sec:roadmap}, we discuss a roadmap for next steps that were beyond the scope of the current investigation. 

%% file: direct-air-capture.tex
\section{Direct Air Carbon Capture and Storage (DACCS) as a basis for PoUW}
\label{sec:direct-air-capture}

In this section, we present a high-level description of direct air carbon capture and storage (DACCS) technology, an argument for why DACCS provides a compelling basis for a PoUW scheme, an outline of the telemetry and mass flow monitoring that can be used in the PoUW protocol, and finally a discussion on the role of life-cycle analysis and net $\text{CO}_2$ removal.

\subsection{Introduction to DACCS}

DACCS is an engineered carbon dioxide removal (CDR) process that involves separating $\text{CO}_2$ from air and then safely and permanently storing it underground. While DACCS is a relatively new technology, there are several scale-up projects underway to demonstrate its feasibility and effectiveness as a climate mitigation strategy. Government support for DACCS is growing, with several countries and regions taking an early lead in supporting its research, demonstration, and deployment. The United States has established several policies and programs to support DACCS, including the 45Q tax credit that provides up to \$180 per ton of $\text{CO}_2$ stored. A key feature of DACCS is that it is not tied to existing energy infrastructure (i.e., fossil-fuel burning power plants), giving it the opportunity to scale by constructing stand-alone DACCS facilities rather than by retrofit. That said, DACCS deployment does rely on the availability of low-carbon energy sources and suitable geological storage. The source of energy used will determine how net-negative the system is and drives overall capture cost, and while the global capacity for storing $\text{CO}_2$ underground is vast, specific site characterization is needed before commissioning a $\text{CO}_2$ storage facility.

The growing voluntary market is the primary driver for investment in DACCS. Large companies that are seeking to deliver on their net-zero commitments can purchase carbon dioxide removal (or a promise for future carbon dioxide removal) to offset their past, current, or future emissions. However, additional incentives for companies to conduct DACCS are needed to drive the technology to the gigaton scale.  

\subsection{Why use DACCS as the basis for PoUW?}

The argument for using DACCS as the basis for a PoUW protocol is based first on its positive climate impact and second on its ability to be precisely measured and monitored. First, a fundamental requirement for the climate-positive blockchain is that its operation results in a net positive climate impact, i.e., $\text{CO}_2$ removal from the atmosphere. The net removal efficiency varies by DACCS technology and energy source, with impacts ranging from –0.36 to –0.94 t$\text{CO}_2e$ per 1t atmospheric $\text{CO}_2$ captured and stored for a baseline grid mix in 2020 \cite{qiuenvironmental}. Second, when compared to nature-based CDR as well as alternative technology-based CDR, DACCS is unique in that its rate of carbon removal can be precisely measured. Using well established measurement tools (e.g., temperatures, pressures, compositions, and flow rates), $\text{CO}_2$ removal rates from DACCS processes can be monitored in real-time. It is this measurement precision that makes DACCS well suited for a carbon negative PoUW scheme.

In our vision, token generation via PoUW can occur in parallel with the voluntary removal purchases to provide additional revenue to DACCS operators, where DACCS operators can sell carbon removal as usual while also using their telemetry to mine a fungible cryptocurrency. If there was widespread adoption of this token as a medium of exchange and store of value, we believe this would greatly accelerate growth in DACCS. 

\subsection{Telemetry and Mass Flow Monitoring}

The goal here is to propose a set of process value measurements (a.k.a., telemetry) that is general across DACCS technologies, that can be used to monitor and verify a facility’s rate of carbon dioxide removal, and that can be communicated to a network to demonstrate PoUW. DACCS involves a capture process coupled to a storage process. The capture rate and storage rate can be measured independently at the points of capture and points of storage, respectively. A process flow diagram and the sensor telemetry that be used to measure and report the $\text{CO}_2$ capture and storage are shown in Figure 1. 

\begin{figure*}[t]
\centering
\includegraphics[width=4.5in]{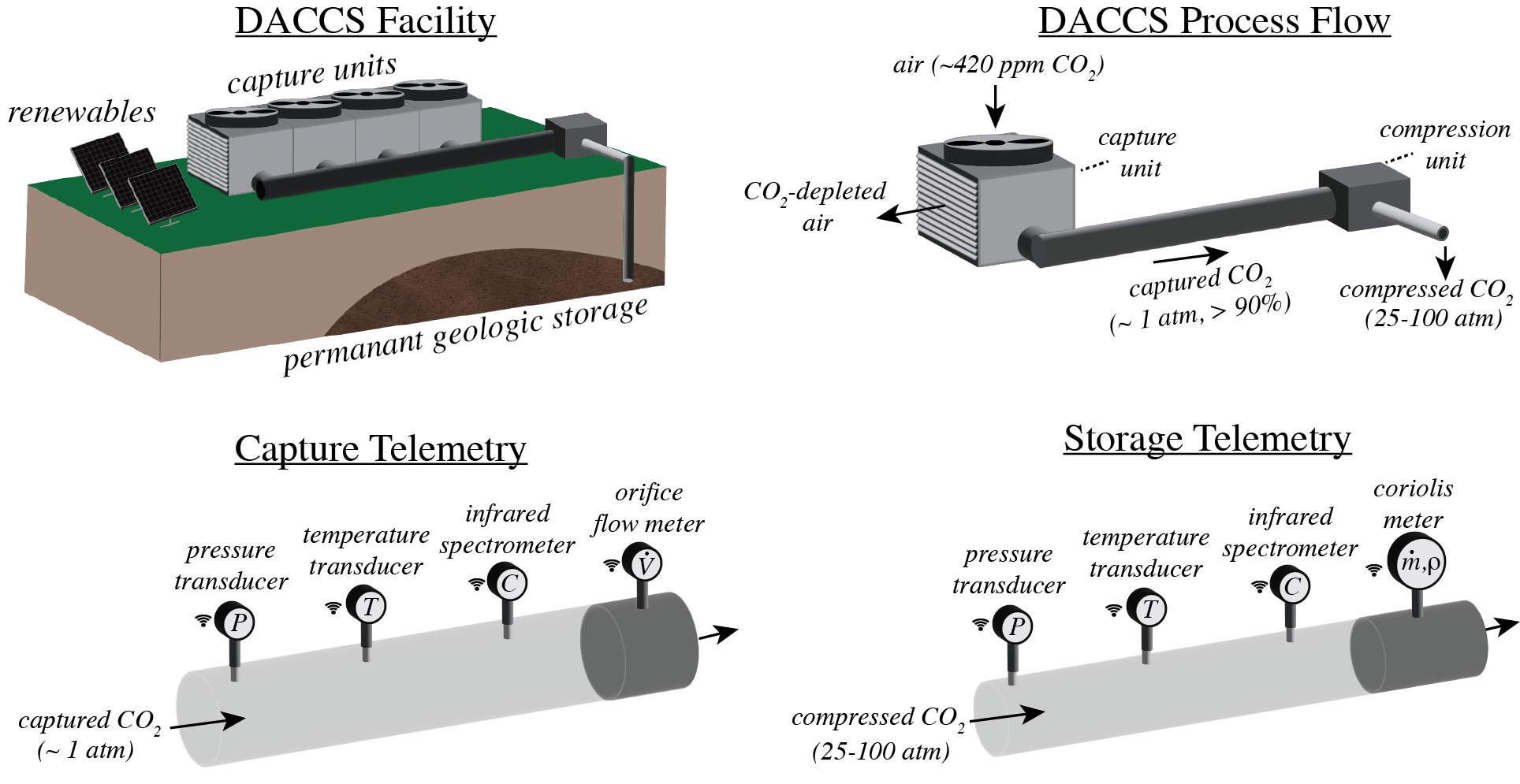}
\caption{DACCS facility, process flow diagram, and sensor telemetry for the capture and storage process.  \label{fig:DACCS Facility and Telemetry}}
\end{figure*}

Here, we propose a monitoring scheme that involves conducting measurements on the capture process and storage process during operation, using those measurements to calculate $\text{CO}_2$ captured and $\text{CO}_2$ stored over a time interval. Capture is generally conducted in a cyclical or continuous process that involves contacting air with a capture media that selectively binds $\text{CO}_2$, regenerating the capture media to produce high purity $\text{CO}_2$ gas, and reusing the capture media in the next cycle. There have been several capture technologies deployed with various types of capture media, wherein the binding can occur via adsorption ($\text{CO}_2$ binding to a solid capture media) or absorption ($\text{CO}_2$ binding to a liquid capture media). Common among capture technologies is an operation in which air is flowed across a capture media, with the concentration of $\text{CO}_2$ in the air decreasing from inlet to outlet. There may be several capture units in parallel operation within the capture facility, so the total capture rate is determined by a sum of the capture rate from each capture unit. The $\text{CO}_{2}$ concentration and mass flow rate can be measured in a variety of ways. For example, the $\text{CO}_{2}$ concentration can be measured using an infrared spectrometer and the mass flow rate can be determined by measuring the volumetric flow rate, temperature, and pressure using an orifice flow meter, temperature transducer, and pressure transducer, respectively.

 After the $\text{CO}_{2}$ is captured, it is compressed and transported to the storage site. The mass flow rate of $\text{CO}_{2}$ can be measured directly using a Coriolis mass flow meter. Alternatively, the mass flow rate can also be measured using a thermal mass flow meter or a turbine mass flow meter. The mass fraction of $\text{CO}_{2}$ can be measured using directly (using an infrad spectormeter) or indirectly. For an indirect measurement of the $\text{CO}_{2}$ mass fraction, the pressure, temperature, and density can be used to infer the mass fraction of $\text{CO}_{2}$ in the stream. The mass of $\text{CO}_{2}$ stored over a time interval at the point of storage can be determined using the mass balance equation:
\[
m_{stored} = \int_{t\textsubscript{0}}^{t} (C_{{CO}_{2}}*\dot{m})dt
\]
\noindent where $C_{CO_{2}}$ is the mass fraction of $\text{CO}_{2}$ and $\dot{m}$ is the mass flow rate. The calculation of net removal from $m_{stored}$ is discussed in the next section.

\subsection{Life Cycle Analysis and Net Removal}

It is important to take into consideration the greenhouse gas emissions associated with constructing and operating the DACCS facility. This can be done using life cycle analysis, where the amount of $\text{CO}_{2}e$ emitted per ton $\text{CO}_{2}$ stored is determined. The life cycle analysis takes into account the energy requirement per ton $\text{CO}_{2}$ stored, the carbon intensity of thermal and electricity sources, the construction and material inputs of the DACCS facility, etc. The life cycle analysis yields a value assigned to the DACCS facility that represents its net removal per ton $\text{CO}_{2}$ stored. DACCS facilities which are the most efficient and use the lowest carbon intensity energy sources will have the greatest net removal per ton stored.  The net removal is given by:
\[
X_{i}=m_{stored}*Y_{i}
\]
\noindent where $X_{i}$ is the net removal of $\text{CO}_{2}$ by a DACCS facility over some time period (ton $\text{CO}_{2}e$), and $Y_{i}$ is the life cycle analysis value assigned to the DACCS facility (ton $\text{CO}_{2}e$/ton $\text{CO}_{2}$). 

%% file: incentive-mech.tex
\section{The Incentive Mechanism}
\label{sec:incentive}

To incentivize DACCS operators, we propose to create a reward mechanism coupled with a mechanism to verify a operator's claims of ${\text{CO}}_2$ capture and storage. We assume that steady state has been reached in terms of the number of DACCS operators. The verification mechanism is necessary to build trust in the reward mechanism, which is necessary to drive widespread adoption and to shift public opinion. There are several ways in which a verification mechanism can be constructed. We do not advocate the use of a trusted entity to drive the verification mechanism, because such a trusted entity is hard to find or develop in practice. 

Instead, in this article, we are concerned with the technological feasibility of a decentralized approach to the verification mechanism, wherein trust is placed in a network of a large number of mutually untrusting validators. No single validator is trusted, however, cryptographic mechanisms combined with verification of sensor telemetry from DACCS facilities can enable a network of validators to verify -- i.e., achieve consensus on -- a DACCS operator's claims of ${\text{CO}}_2$ removal.

Censorship resistance implies that anybody on the planet can set up a DACCS facility and benefit from the associated reward mechanism. This would require that the technical specifications of a prototypical DACCS facility and its operating procedures be available to everybody. This is not unlike the censorship resistance achieved in cryptocurrencies such as bitcoin.

In the subsequent sections, we will describe how to implement this combination of a reward mechanism coupled with a verification mechanism. For concreteness, the CDR operators in these protocols are DACCS facilities, but extensions to other approaches are possible. 
We describe a winner selection protocol, in which every DACCS facility performs the ``useful work'' of ${\text{CO}}_2$ removal and wins a reward with a probability proportional to the amount of ${\text{CO}}_2$ it has (provably) removed. 

When the reward is in the form of a crypto token, we describe how the winner selection protocol lends itself to a blockchain-based cryptocurrency based on PoUW. The verification mechanism in this case, thus establishes consensus not only on which DACCS facility wins the reward for ${\text{CO}}_2$ storage, but also on the state of the resulting blockchain. We remark that, unlike popular PoW mechanisms such as bitcoin, the ``work'' done by \emph{all} DACCS facilities (not just the winning DACCS facility) is climate positive. We contend that, if this token is widely adopted as a medium of exchange, it would provide an additional powerful incentive for ${\text{CO}}_2$ removal. In other words, the token would not only support CDR activities but economic activity more broadly and would potentially become a medium of exchange.  

In addition, our proposed incentive mechanism takes advantage of a key benefit to issuers of most fiat currencies, such as the US dollar, called seigniorage. The actual production of these tokens is close to zero but the token represents monetary value.\footnote{For discussion of seigniorage, see \cite{Haslag98seigniorage}.} The difference between the face value of the token minus the production cost is known as seigniorage. As the demand for CDR removal increases, so does the monetary value of the token. Thus, the increase in demand for the coin creates additional monetary incentive to DACCS operators.

%% file: consensus-mech.tex
\section{Designing A Consensus Mechanism and Reasoning about Security Guarantees}
\label{sec:consensus}

A key technical challenge is to craft a distributed consensus protocol in which mutually untrusting parties can verify claims of $\text{CO}_2$ captured and stored, and for each round, determine a winner based on the amount of $\text{CO}_2$ captured and stored. The role of such a winner can be many-faceted. For example, the process of determining a winner may trigger the generation of a specified quantity of crypto tokens which are sent to the winner. The winner -- in this case, the winning Direct Air Carbon Capture and Storage (DACCS) facility -- could, in turn, distribute the token amongst several stakeholders, use the token, directly or converting it to medium of exchange, to fund the DACCS operation, or use it to purchase other goods and services. In a cryptocurrency application in particular, the winner records the awarded tokens in the block they create and append to the blockchain. 
We provide below the key steps in a notional protocol for establishing consensus about the DACCS facility that receives a reward for storing ${\text{CO}}_2$  in a climate-positive blockchain.

\noindent
\subsection{Participants} 
The consensus protocol is executed by a network of mutually untrusting entities. A subset $\mathcal{M}$ of these entities own or operate carbon capture facilities. For specificity, we will assume that the ${\text{CO}}_2$ capture facilities employ DACCS technology. We will refer to the DACCS facilities as miners in our context for two reasons: (1) they are extracting ${\text{CO}}_2$ from the air and storing it, (2) their role is analogous to miners in traditional cryptocurrency mining~\cite{nakamoto08bitcoin}. 

Let the set of miners be denoted by $\mathcal
{M} = \{M_1, M_2, \ldots, M_K\}$. Another subset $\mathcal{V}$ of the mutually untrusting entities is tasked with receiving information from the DACCS facilities, verifying the claims of carbon capture and storage, and determining the winning DACCS facility which earns a reward in the form of a token. We denote the set of these validators as $\mathcal{V} = \{V_1, V_2, \ldots, V_N\}$. Note that $N \gg K$, as any one with a relatively small amount of computing resources can sign up to be a validator, while setting up a DACCS facility is an expensive proposition. In some variations, there can be overlaps between the two sets, i.e., a DACCS facility can also be a validator, thus $V_i \equiv M_j$ for some $i < N, j < M$. For the development below, however, we consider the simpler case in which the two sets to be disjoint.

\noindent
\subsection{Winner Selection Protocol} 

\begin{figure*}[t]
\centering
\includegraphics[width=5.0in]{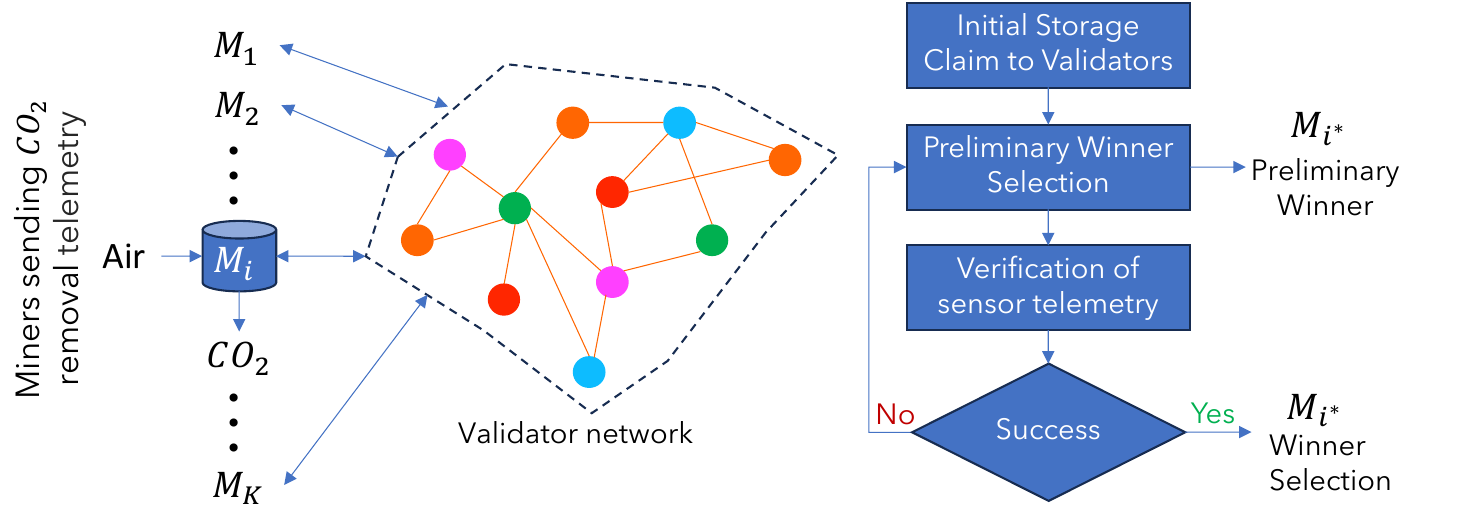}
\caption{DACCS operators submit sensor telemetry to a network of validators which determines a winning DACCS operator (miner). \label{fig:winner-selection}}
\end{figure*}

We now describe how a network of untrusted miners and untrusted validators can establish consensus about the amount of $\text{CO}_2$ captured and stored by one of the miners chosen at random (See Fig.~\ref{fig:winner-selection}). If the $\text{CO}_2$-capture and storage claims are verified, the chosen miner is determined the winner, and a reward -- in terms of a quantity of crypto tokens --  is generated and sent to the winner. We defer a discussion of the value of the token and its possible uses to subsequent sections. Winner selection occurs using the following steps:
\begin{enumerate}
    \item \textbf{$\text{CO}_2$ Capture:} At a designated time-stamp $t_s$ agreed upon by the miners and validators, each miner $M_i \in \mathcal{M}$ starts capturing $\text{CO}_2$. The capture process continues for a designated constant period of time $T$. The miner stores the sensor data from the capture process. Let us denote the sensor data captured by $M_i$ as $\mathcal{C}_i(t) = \{C_i^{(1)}(t), C_i^{(2)}(t), \ldots, C_i^{(P)(t)} \}$, where $t_s \leq t \leq t_s + T$. Here, $C_i^{(1)}(t), C_i^{(2)}(t),$ etc, are the various physical variables that constitute the $\text{CO}_2$-capture process. 
    \item \textbf{$\text{CO}_2$ Storage:} At regular intervals after time $t_s$, each miner also performs a process of storing the captured $\text{CO}_2$. The cadence of capture and storage intervals can differ from miner to miner, depending upon the equipment and technology used. The storage process can be performed in chunks of time, or can be performed continuously as more and more ${\text{CO}}_2$ is captured. The miner stores the sensor data from the storage process. Let us denote the sensor data captured by $M_i$ as $\mathcal{S}_i(t) = \{S_i^{(1)}(t), S_i^{(2)}(t), \ldots, S_i^{(Q)}(t) \}$, where $t_s \leq t \leq t_s + T$. Here, $S_i^{(1)}(t), S_i^{(2)}(t),$ etc, are the various physical variables that constitute the ${\text{CO}}_2$-storage process. The sequence of ${\text{CO}}_2$ capture and storage processes, shown here in the context of direct air capture continue until a time stamp $t_e = t_s + T$.  
    \item \textbf{Initial ${\text{CO}}_2$-Storage Claim to Validators:} Each miner $M_i, 1 \leq i \leq M$ then sends to each validator $V_j, 1 \leq j \leq N$, an initial claim of ${\text{CO}}_2$ storage. This communication from $M_i$, received by each $V_j$ is described as a set $\mathcal{A}_i = \{i, r_i, t_s, X_i, \text{ZKP}(\mathcal{C}_i(t), \mathcal{S}_i(t)) \}$. Here, $\text{ZKP}(\mathcal{C}_i(t), \mathcal{S}_i(t))$ is a concise non-interactive zero-knowledge proof extracted from the capture and storage telemetry.\footnote{The zero-knowledge 
    (ZK) proof captures the physical relationships between the sensor readings in the capture and storage processes. We are not aware of such proofs existing today, but are of the opinion that succinct ZK proof approaches developed in the literature (e.g., \cite{bitansky12tcs,ben18eprint}) could be leveraged to construct them. We pose the development of succinct Non-Interactive Zero-Knowledge Proofs (NIZKs) to capture relationships amongst physical quantities as an interesting and open research area.} $X_i$ is the claimed amount of stored $\text{CO}_2$. $r_i$ is a random number that will later be used by the validators for selecting a winning miner. To ensure message authenticity and integrity, $\mathcal{A}_i$ is bundled with a digitally signed hash, computed by $M_i$, given by $\text{Sign}_i(h(\mathcal{A}_i))$, where $h(\cdot)$ is a standard hash function, e.g., SHA-256. The final message thus takes the form $ \mathcal{B}_i = \{ \mathcal{A}_i,\text{Sign}_i(h(\mathcal{A}_i))\}$.
    \item \textbf{Verification of Initial Claim:} Each validator $V_j$, then verifies that the message received from each $M_i$ is genuine and unmodified, using the public key of $M_i$ and the knowledge of the hash function $h(\cdot)$. Each validator also verifies the zero-knowledge proof $\text{ZKP}(\mathcal{C}_i(t), \mathcal{S}_i(t))$ for each miner. This means that each miner presents a proof to the validators that they have captured relevant sensor data during capture and storage, without revealing that data. At this point, a number of mechanisms are possible to enable the validator network to agree upon a winning miner. We describe one possible mechanism here, but many others are permissible. 
    \item \textbf{Preliminary Winner Selection:}
    To each miner $M_i$, a validator then assigns the length of a subset of the unit interval $[0,1]$ given by:
    \begin{equation}
        \ell_i = \frac{X_i}{\Sigma_i^M X_i} \label{eq:interval}
    \end{equation}
    To maintain consistency amongst the calculations of the validators, the calculation of $\ell_i$ is performed by a validator $V_j$ if and only if they have received the message $\mathcal{B}_i$ for each $i \in \{1, 2, \ldots, M\}$. If not, the validator does not participate in this computation. Furthermore, the segments of length $\ell_i$ are arranged in ascending order of $i$, such that $\Sigma_i^M \ell_i = 1$. Let the $i^{\text{th}}$ segment, having length $\ell_i$ have endpoints $u_{i-1}$ and $u_i$. Thus, the lengths are given by $\text{len}(u_0, u_1) = \ell_1$, $\text{len}(u_1, u_2) = \ell_2$, $\text{len}(u_0, u_2) = \ell_1 + \ell_2$, $\text{len}(u_0, u_M) = \Sigma_i^M \ell_i = 1$, etc.  Each validator $V_j$ computes a random number, which -- by construction -- is the same across all validators, given by:
    \begin{equation}
        \bar{r} = \sum_{i=1}^{M} r_i \notag    
    \end{equation}
    The number $\bar{r}$ is then hashed using a cryptographic hash algorithm, e.g., SHA-256, and mapped to the unit interval as a number $\widehat{r} \in [0,1]$. Then each $V_j$ determines the winning miner as:
    \begin{equation}
        M_{i^*} \equiv M_i\,\,\,\text{such that}\,\ \text{len}(u_0,u_{i-1}) \leq \widehat{r} \leq \text{len}(u_0, u_i) 
    \end{equation}
    Each $V_j$ broadcasts or gossips the identity of the winning miner $M_{i^*}$ to the validator network, along with the random number $\widehat{r}$ so that other validators can verify that the preliminary winning miner is chosen correctly. Concretely, $M_{i^*}$ is considered to be declared as the preliminary winning miner when the number of validators confirming the choice of $i^*$ crosses a specified (large) threshold. 
    \item \textbf{Verification of Sensor Telemetry}: When the preliminary winning miner $M_{i^*}$ is determined, its identity is broadcast to the network. $M_{i^*}$ then transmits its actual sensor data, i.e., $\mathcal{C}_{i^*}(t)$ and $\mathcal{S}_{i^*}(t)$ to each validator $V_j$. In addition to the raw sensor data, $M_i*$ also sends to the validators digitally signed hashes of the capture and storage data, accompanied by the timestamp $t_s$. We denote this signed, timestamped data by $\text{Sign}(t_s \|  h(C^{(k)}_{i*}(t_s) \| C^{(k)}_{i*}(t_s+1) \| \ldots \| C^{(k)}_{i*}(t_s + T))$ for $1 \leq k \leq P$ on the capture side, and $\text{Sign}(t_s \| h(S^{(m)}_{i*}(t_s) \| S^{(m)}_{i*}(t_s+1) \| \ldots \| S^{(m)}_{i*}(t_s + T))$ for $1 \leq m \leq Q$ on the storage side. Here again, $h(\cdot)$ is a standard hash function. What this means is that each sensor on the capture and storage side digitally signs its telemetry and incorporates a timestamp.
    Each validator then confirms whether the capture and storage data are consistent with $X_{i^*}$, the claimed amount of $\text{CO}_2$ captured.\footnote{We remark that an alternative approach would refrain from splitting the ${\text{CO}}_2$ storage claims into a preliminary check and a final check. Concretely, if the zero-knowledge proof can be designed to capture the relationship between $X_i$ and the capture/storage sensor readings, then the last step of the protocol is not necessary. Again, development of efficient zero-knowledge proofs to accomplish this is posed as an open research problem.} If the sensor telemetry is determined to be consistent by each validator, (or a large enough majority of validators), then the winning miner $M_{i^*}$ is confirmed. If not, the step of preliminary winner selection is repeated after excluding $M_{i^*}$, i.e., after setting $X_{i^*} = 0$ and therefore $\ell_{i^*} = \text{len}(u_{i^*-1},u_{i^*}) = 0$. This process is repeated until it is confirmed that the sensor telemetry received from the chosen miner is consistent with the claimed amount of carbon storage for $t_s < t < t_s + T$.
 \end{enumerate}

\subsection{Security and Privacy Considerations}
\label{subsec:security-winner-selection}
We presented the above protocol with clarity in mind, however, there are alternative implementations that might be more efficient from the perspective of the protocol's communication overhead and privacy of the miners. As an example, it is conceivable that the step representing verification of sensor telemetry in Step 6 might be obviated by absorbing the verification operations inside a novel zero-knowledge proof in the step representing verification of the initial claim (Step 4). We do not yet know of an efficient zero-knowledge proof mechanism for this alternative implementation. Conceptually, it appears to be similar to zero-knowledge middleware boxes (ZKMB)~\cite{grubbs22usenix} used to prove compliance with network protocol requirements. In that case, the winner selection process would primarily be performed in Step 5, with just the mitigations for non-compliant telemetry -- i.e., disqualifying miners whose telemetry has been shown in zero-knowledge in Step 5 to be inconsistent with the claimed ${\text{CO}}_2$ storage values -- being performed in Step 6. In what follows, we will discuss security considerations that apply to the protocol as described above, wherein the winner selection is done in two steps, a preliminary selection (Step 5) followed by a verification of sensor telemetry (Step 6). Security considerations include:
\begin{enumerate}
    \item \textbf{Data Authenticity and Integrity:} The protocol described above ensures authenticity and integrity of the information being transmitted. In the initial storage claim, each miner $M_i, 1 \leq i \leq K$ digitally signs the data sent to the validators. This prevents man-in-the-middle attacks in which the data is intercepted and manipulated in transit, before it reaches any validator. Furthermore, in the verification of the sensor telemetry, the data sent from the capture sensors and the storage sensors to the validator network is also digitally signed by the sensor manufacturer. We assume that the sensors are equipped with a trusted platform module at the time of manufacture, that allows them to sign their data. The signature verification keys of all the sensors are publicly available to all validators. This ensures that the miner operating the DACCS facility cannot modify the sensor data to claim a false amount of $\text{CO}_2$ captured or stored. Concretely, a dishonest miner could fabricate sensor data but would not be able to produce a digital signature on the fabricated data because he does not know the sensor's signing key.
    \item \textbf{Replay Attacks:} A timestamp is embedded in the signed data sent by each capture and storage sensor. This ensures that a miner cannot replay sensor data from the past while fooling the network that it has newly captured or stored $\text{CO}_2$.
    \item \textbf{Faulty Initial Storage Claim Data:} Suppose that a validator is not able to verify the zero-knowledge proof that a particular miner $M_i$ has gathered relevant sensor data. Note that the protocol requires the same ZKP to be sent to all validators. If some validators discover that they are unable to verify the ZKP sent by $M_i$, then it means that $M_i$ has sent a different and incorrect ZKP to some or all validators, so this miner must be disqualified from the winner selection algorithm. In this case, the validator's gossip the identity of the miner $M_i$ and exclude that miner from the further steps in the protocol. This is essential because all validators have to maintain the same state -- the same random number $\bar{r}$ and the same segments of the unit interval in the preliminary winner selections step -- in order to compute the preliminary winner. 
    \item \textbf{Sensor Data Incompatible with Storage Claim:} Suppose that, in the final step involving verification of the sensor telemetry, one or more validators discover that the amount of $\text{CO}_2$ claimed as stored by $M_{i^*}$ is inconsistent with the received sensor data. In this case, at a minimum, $M_{i^*}$ is removed from consideration, i.e., the protocol repeats step 5 after setting $X_{i*} = 0$. It is conceivable that stricter measures may be adopted against $M_{i^*}$ to discourage dishonest $\text{CO}_2$ storage claims by miners, such as disqualification from participation for a longer time period; We do not consider such measures in detail in this paper.
    \item \textbf{Validators that Intentionally Misreport Results of one or more Protocol Steps:} Due to the use of digital signatures, it is computationally infeasible for a validator to forge the sensor telemetry in both the verification of the initial claim of carbon capture, and the verification of the full sensor data. Malicious action, in this case, consists of a collusion of a vast number of validators who coordinate to misreport the result of the preliminary winner selection algorithm. In essence, this is similar to a malicious attack on a cryptocurrency where a majority of the nodes refuse to confirm a block transaction, for the purpose of generating a temporary fork in the blockchain. The larger the size of the set of colluders, the more secure the protocol is from their actions. At the very least, this number should be set to be greater than 50\% of the total number of validators.
\end{enumerate}

\subsection{Complexity Considerations}
The computational complexity incurred at each DACCS facility is dominated by the computation of digital signatures and  zero-knowledge proofs in Step 3. The complexity incurred at each validator is dominated by the verification of the zero-knowledge proofs in Step 4, and the verification of the sensor telemetry in Step 6. Taken together, this complexity is minuscule compared to that incurred in conventional PoW mechanisms. For e.g., in bitcoin, a hash function, such as SHA256, has to be computed billions of times (in general) until the output satisfies a stipulated mathematical condition. Since these repeated operations have been replaced by ${\text{CO}}_2$ capture and storage in our approach, the computational complexity of the winner selection protocol is significantly lower than that of traditional PoW mechanisms.  

\subsection{Fault Tolerance Considerations (Missing Initial Storage Claim Data)}
    Suppose a validator does not receive the initial storage data from some miner $M_i$. Then, this validator cannot implement the preliminary winner selection step. We stipulate that any validator that has missing storage data does not participate further in this round of establishing consensus, i.e., the round corresponding to the time interval $[t_s, t_s+T]$. A key design parameter is the number of validators that must have all the storage claim data presented by the miners, and must stay online during the entirety of the protocol execution. This number should be large enough, otherwise the winner selection mechanism would not be credible.

\section{Leveraging the Consensus Mechanism in a Blockchain-based Cryptocurrency}
\label{sec:leveraging}

A key motivation for the winner selection protocol is to develop mechanisms that incentivize $\text{CO}_2$ capture and storage. This motivation leads to a reward mechanism and an associated cryptocurrency in a natural way as we describe below. For the purposes of this article, a cryptocurrency has the additional property versus a token that it can be used in the broader economy to buy and sell goods and services. To clarify our development below, we will often refer to the architectural components of Bitcoin~\cite{nakamoto08bitcoin}, the world's most popular cryptocurrency, by way of analogy.

\subsection{A Reward in Crypto Tokens}
Suppose that the winner selection algorithm of the previous section is executed, and a winning miner $M_{i^*}$ is determined. Suppose that, as a result of being the winner, $M_{i^*}$ is rewarded with a quantity $Z$ of crypto tokens. In other words, the conclusion of the winner selection algorithm results in the creation of $Z$ new tokens. The winner, $M_{i^*}$ may then convert the token into a medium of exchange to purchase goods and services. Alternatively, the reward could be saved for future use.

For the cryptocurrency application, we remark that the identity of the winning miner $M_{i^*}$ need not be revealed. This can be accomplished by using an identifier, such as the miner's public key $PK_i$ rather than the index $i$. For the winner selection algorithm of Section~\ref{sec:consensus} to work with such an identifier, the steps of that protocol can all be executed by replacing $i$ with $PK_i$, with one exception: The preliminary winner selection step needs an explicit ordering amongst the identifiers of the miners, to create the partitions of the unit interval using which the winning miner is randomly chosen. With the public key identifiers, we no longer have the naturally ordered identifiers in a set $\{1, 2, ..., M\}$. However, we can obtain an ordering of $M$ elements in many ways, the simplest being to order the identifiers $PK_i, i \in {1, 2, ..., M}$ in increasing order of magnitude such that $j < i \implies f(PK_j) < f(PK_i)$, where $f$ is a suitable function. We assume that such an $f$ always exists. With the new ordering, it is again possible to generate a partition of the unit interval and choose a uniformly distributed random variable over the unit interval to choose the winning miner. In this case, since the index of the winning miner maps back to a public key identifier rather than a miner's identity, it preserves the anonymity - though not the \emph{privacy} - of the miner. Anonymity can be improved if the miners use new public-private key pairs in each round, and for each transaction that they engage in, thus making it difficult to link successive transactions or reward accruals.  

\subsection{Crypto-Token Transactions in the Real World}

When a crypto token is used to purchase goods and services in the economy, the token serves as a medium of exchange. The total supply of a cryptocurrency can grow indefinitely such as for ether or be capped such as bitcoin. In addition, the supply of cryptocurrencies can also decrease by permanently removing coins from circulation. In this article, we abstract from determining the optimal supply of the new token based on the demand for the token. As noted before, although, the new crypto token is produced based on competition among DACCS facilities, the value of the token is not connected to the amount of $\text{CO}_2$ captured and stored.  

We will now describe an approach in which transactions involving such a token are recorded in a blockchain that is driven by the climate-positive consensus protocol from Section~\ref{sec:consensus}. Consider that Alice and Bob engage in a transaction in which Alice must pay $z$ tokens to Bob. Assume that Alice and Bob possess digital wallets similar to the ones available for cryptocurrencies today, and that they are each identified using their public keys. The transaction can then be reduced to the following quantities (1) the input address, i.e., Alice's public key, from which the token is sourced, (2) the amount $z$ of the transaction, (3) the output address, i.e., Bob's public key, to which the token will be transferred when the transaction is complete, (4) optionally, an amount $\Delta z$, consisting of the transaction fees that may be paid to the miner, (5) a time stamp of the transaction.

\subsection{Protocol for Incorporating Transactions in the Blockchain}

\begin{figure*}
\centering
\includegraphics[width=5.5in]{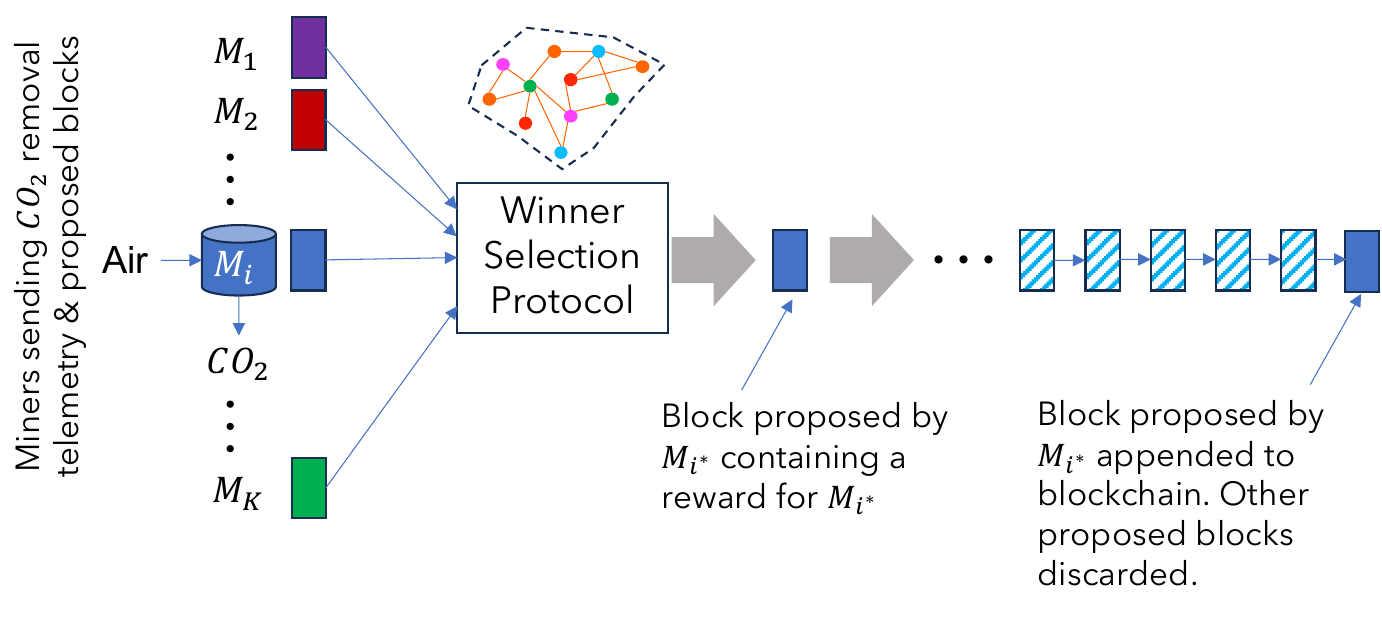}
\caption{The DACCS operators can leverage the winner selection protocol to propose candidate blocks of crypto-token transactions that are appended to a blockchain. \label{fig:blockchain-protocol}}
\end{figure*}
The protocol for transferring $z$ tokens from Alice to Bob, and recording that transaction in the blockchain would then follow the steps below (See Fig.~\ref{fig:blockchain-protocol}):

\begin{enumerate}
    \item \textbf{Creation:} Bob sends his public key $PK_B$ to Alice. Alice then creates the transaction file,and signs it with her private key $SK_A$. Thus, the transaction involves sending $z + \Delta z$ tokens from input address $PK_A$ to output address $PK_B$ at some time $t$.
    \item Alice sends the signed transaction to the closest node in the blockchain, which could be a miner or a validator. From there, it is propagated to the validator network using a suitable gossip protocol~\cite{birman07sigops}.
    \item \textbf{Preliminary Verification:} The transaction undergoes preliminary verification in which basic requirements are checked. These include, for example, checking that the input address has a balance greater than $z + \Delta z$ tokens at the time stamp $t$. Once the transaction clears the preliminary verification, it is transferred to a pool, analogous to Bitcoin's Memory Pool, where it awaits a miner.
    \item Any miner $M_i \in \mathcal{M}$ can then consider the transaction for inclusion in the next block of transactions to be mined. We will focus on the activity of a single miner $M_i$, one of possibly many that will include the transaction in a candidate block. One criteria for selecting the transaction in the proposed new block is whether the transaction includes a fee intended for the miner. 
    \item Each $M_i$ chooses a specified number, $\eta$ of new verified transactions for inclusion in a candidate block that he will propose to the network. One approach for $M_i$ to choose transactions from the Memory Pool for inclusion into the new block is to select those with the highest transaction fee. The miner also adds the following information to the proposed block:
    \begin{enumerate}
        \item A cryptographic hash of the most recent block added to the blockchain. This can be a SHA-256 hash, and it cryptographically chains the blocks.
        \item A Merkle tree~\cite{merkle80sp} computed from the hashes of all the transactions included in the block.
        \item The amount of ${\text{CO}}_2$ stored during the current time block, indexed by $t_s$, in the winner selection protocol. This would be indicated by $X_{PK_i}$, rather than $X_i$, as used in the winner selection protocol.
    \end{enumerate}
    This additional information allows anyone inspecting the block to determine how much ${\text{CO}}_2$ was stored by the winning  miner. Note that an aggregate of the stored amounts for all blocks is significantly lower than the total amount of ${\text{CO}}_2$ stored by the blockchain, because the miners that were not chosen as winners have also performed useful work by contributing to ${\text{CO}_2}$ stored. This is a key beneficial aspect of the climate-positive blockchain\footnote{Since the $
    {\text{CO}}_2$ storage amounts are gossiped by the DACCS facilities to the network, the protocol may be modified to include the total amount of ${\text{CO}}_2$ stored by $all$ miners while mining the \emph{previous} block in the blockchain. With this modification, aggregating the total ${\text{CO}}_2$ storage amounts over the entire blockchain provides an estimate of the ${\text{CO}}_2$ stored since the deployment of the blockchain. The caveat is that, by construction, only the ${\text{CO}}_2$ storage amount of the winning miner has been verified for each block; the storage amounts for all other miners are unverified, and therefore should only be considered as estimates.}. 
    \item Each $M_i$ then includes a special transaction in which the miner pays into its own digital wallet an amount given by
    \begin{equation}
        \theta = Z + \sum_{j}^{\eta} \Delta z_j \notag
    \end{equation}
    where $\Delta z_j$ is the transaction fee in the $j^{\text{th}}$ transaction included in the candidate block. The value $\theta$ is thus the sum of the mining reward and any transaction fees. This is similar to bitcoin's \emph{coinbase transaction}, in which new crypto tokens are created. Also, similar to bitcoin, this transaction is special in the sense that the total earnings $\theta$ cannot be spent unless a specified (large) number of vaildators have confirmed that candidate block has been appended to the blockchain, thereby confirming that $M_i$ was the winner of that round, i.e., ($i\equiv i^*$). If, on the other hand, $M_i$ is not chosen as the winner in the winner selection algorithm, the new proposed block of transactions (along with the coinbase transaction) is discarded. $M_i$ will propose a new block in the next round.  
\end{enumerate}
The reader will note that the structure and construction of the blocks resembles that in Bitcoin. The difference lies in the fact that, in Bitcoin, the block records the block hash as well as nonce that the winning miner used to successfully mine the block, while in the proposed climate-positive crypto-token, the block records the amount of  ${\text{CO}_2}$ stored by the miner and confirmed by the validator network. 

\subsection{Security and Privacy Considerations}

Many of the security considerations for this protocol are inherited from those in the winner selection protocol, as described in Section~\ref{subsec:security-winner-selection}.

\begin{enumerate}
\item \textbf{Protocol Security}
We do not claim that the protocol described here is as secure as bitcoin. This is because of a fundamental difference: The Bitcoin protocol operates entirely in the digital world and is grounded in well-understood cryptographic mechanisms that have stood the test of time. Our protocol, on the other hand, attempts to connect a physical process (${\text{CO}}_2$ removal) to a digital protocol with cryptographic mechanisms. Thus, attacks on the proposed protocols can come from both the physical and the digital worlds, and crucially, at the interface of the physical and digital worlds.\

Security mechanisms for preventing subversion of the physical process are a work in progress. We have described how signing the sensor telemetry can help avoid security problems related to data authenticity and integrity. Incorporating a time-stamp in the digital signature is used to deter replay attacks. We anticipate the need to develop further efficient mechanisms (both physical and digital) to improve the security of the physical process, as well the security of its interface with the digital process. We hope that this work will highlight the need and catalyze further research in the field.
\item \textbf{Anonymity of Miners}: The protocol, as described, has one key difference compared to consensus mechanisms based on a cryptographically hard problem. In our protocol, the validator network knows the public key of the miner when it executes the winner selection protocol. This is slightly different from Bitcoin, in the sense that (1) the miners who propose a candidate block in the Bitcoin blockchain can do so without easily revealing their public key (2) the public key of the winning miner can only be accessed by someone who can read the scriptsig field of Bitcoin's coinbase transaction, and then the miner can possibly be identified by analyzing transactions on mining pools. In other words, it is easier to identify the public key of the winning miner in our proposed mining process, compared to the case in bitcoin. As stated earlier, if the miners refresh their public/private key pairs for each new mining round, they achieve better anonymity. Like Bitcoin, however, their identities could still be revealed via analytics on blockchain transactions. This difference has implications for oversight into the operation of the DACCS operators (miners). The cryptographic machinery of Bitcoin ensures that it needs no oversight beyond verifying the hash computations. For CDR technologies, the scope of oversight is broader. While we verify sensor telemetry and ensure data authenticity and integrity in our protocols by means of cryptographic tools -- thus making it difficult for a rogue DACCS operator to fake telemetry -- it may be necessary to supervise other aspects of the DACCS operation. For example, is the storage being performed in a geologically suitable region using the correct processes so that the stored ${\text{CO}}_2$ does not escape back into the atmosphere? Are the storage readings taken at the correct stage in the process? It would be useful to perform such checks at random, sparse intervals. Who should perform this oversight? These are open questions that need to be addressed. 
\item \textbf{Anonymity of Transacting Entities:} We remark that the level of anonymity provided to entities that engage in actual transactions of our proposed token, is identical to that in Bitcoin. That is because the transacting entities are only involved in the digital components of consensus mechanism, and have no involvement in the verification of sensor telemetry.
\end{enumerate}

 

%% file: roadmap.tex
\section{Roadmap: scaling needs, generalizability and open questions}
\label{sec:roadmap}

While novel frameworks such as the one presented here, need a pathway for adoption and scaling, the technologies behind the platform will also keep improving. In this section, we deliberate on four broad areas: (a) the needs for scaling such a framework, (b) discussion of tokenomics (c) the generalizability of such a framework to other carbon capture technologies, and (d) some open unanswered questions.

\subsection{Scaling needs of the framework}

The proposed solution considers the steady state situation, where there are sufficient number of DACCS to capture carbon and likely enough ways of storing or upconverting the captured carbon. However, to scale up from the first established DACCS to a steady state condition of $\mathcal{M}$ miners, there needs to be a suite of incentive mechanisms in place for a number of participants in the ecosystem. The following can be viewed as points aiding the scaling of the framework or also as bottlenecks which might limit scaling possibilities:
\begin{enumerate}
    \item \textbf{DACCS:} The ability of a miner to effectively use the awarded crypto-tokens needs the token to be liquid. This need includes a mechanism to trade it for other forms of hard or cryptocurrency and the ability to handle cross-currency swap risks. If the tokens display liquidity, adoption of the framework to newer DACCS will be much faster. Establishing a tokenomics model of the crypto-tokens will be a useful exercise in promoting the liquidity.
    \item \textbf{Local communities:} Setting up a DACCS needs land, which needs to be procured, often from local communities already established on the land. Although a general concern with DACCS, with an increased monetary value associated with the facilities as brought about by our framework, there may be several repercussions. The price of land may go up giving rise to two possible outcomes. Firstly, it may make establishing new DACCS less attractive given the returns (which may lead to more vertical facilities) thereby driving lower returns for the farms and local community. Alternately, the price rise may help local communities and create more efficient DACCS, all while increasing the amount of captured carbon. Noise pollution and the potential long term safety of stored underground carbon may prove to be a repellent, similar to wind farms. A secondary market can provide additional leverage and participation in accelerating adoption of the framework.
    \item \textbf{Technology companies:} It can be expected that there will be frequent improvements in DACCS technologies and components including in sorbents, sensors, sequestration and upconversion processes. Being a long term investment for DACCS, it may come to pass that such technology companies are under beholden contracts with the facilities to ensure a steady dedicated supply of materials and component devices. To make the sensors used along various points of the process steps tamper-proof, they may need unique certification by the manufacturers. This may enable longer term contracts between DACCS and OEMs, thereby making it attractive to both parties.
    \item \textbf{Local governments:} Any modification of the local environment by carbon capture opens the chance of inadvertent minor engineering of the local climate. This may need the open discussion between local, state and federal government bodies. If the DACCS are established close to the boundary between two neighboring countries, it may result in the need for a greater negotiation between multiple parties. If done well, this can have positive changes to local environments thereby reflective on local economies.  
\end{enumerate}

\subsection{Tokenomics}

Tokenomics refers to the underlying economics of tokens including factors that determine their optimal token supply based on the token demand. Although we do not discuss price dynamics of the token in detail, we can offer some insights. First, as the demand for CDR increases, the price of the token should rise given no or little change in supply. Second, as the DACCS technology improves, the price of the token falls given little change in demand. Of course, there could be increases in both supply and demand simultaneously.  

Furthermore, decisions regarding whether the total supply of the token will be fixed or unbounded need to be explored. Discussions about the path of optimal supply of the token over time is beyond the scope of this article. In other words, are there rules on how supply increases or decreases? Does the community determine when to burn or take tokens out of supply? 

Initially, the demand of the token will be based on the trust of those that want to participate in DACCS technology for CDR. What alternatives do these participants have and how does this token rank among those alternatives? If successful, the demand for this token along with its price will also increase. Once the token is established, it may be used for a host of other transactions. What token characteristics would be desirable for other types of transactions? Clearly, the positive impact on the environment resulting from the PoUW would be extremely attractive to users.

Also, as the price rises, these tokens may circulate less because holders of these coins prefer to hold on to them instead of transacting with them leading to reduced liquidity. Decentralized finance platforms may evolve to increase liqudity where holders would deposit their tokens to liquidity pools. These discussions are beyond the scope this article. 

As new CDR technologies are adopted, how would DACCS tokens interact with other types of tokens. Would there be a market-based conversion rate between different types of CDR tokens?

\subsection{Generalizability of the Framework}

Thus far, the framework has been described with the DACCS as an example. However, it can be extended to many other forms of climate positive actions, not restricted to carbon capture. This includes ocean carbon capture, DACCS + S (DACCS and sequestration), DACCS + U (DACCS and utilization/upconversion), capture of other greenhouse gases such as methane ($CH_{4}$), nitrous oxide ($N_{2}O$), (with present or future sorbent materials), and ocean de-acidification, to name a few. Any climate positive action thus, involving the capture, production or modification of a known parameter along with a relevant set of process steps having periodic measurements with multiple sensors can be amenable to this protocol/framework. 

\subsection{Open Questions}

While goodwill is an underlying ethos behind the framework, for improving the future of humanity on earth, it also provides a financially lucrative reward for doing good. Furthermore, since the consensus mechanism that drives the reward is achieved algorithmically, it is essential that technological means be developed to ensure correctness and security of the protocols.\\

\subsubsection{\textbf{Technological Considerations}}
\begin{enumerate}
\renewcommand{\labelenumi}{\alph{enumi}.}
    \item \textbf{Sensor Design and Manufacturing:} The protocols we describe require sensors to digitally sign the data they produce, because this preserves authenticity and integrity of the sensor data. To achieve this capability, it is necessary to augment current sensor technology -- for example with hardware enhancements such as Trusted Platform Modules (TPMs)~\cite{kinney06book} -- to enable them to sign the data. This would increase the cost of the sensors, and it remains to be seen whether the scale at which such sensors are manufactured will make the cost manageable in the long run.
    \item \textbf{Protocol Design:} As we have described, zero-knowledge proofs can enable DACCS facilities to efficiently and securely verify their sensor telemetry and their ${\text{CO}}_2$ storage claims. We did not, however, provide an implementation of the proofs. Developing efficient non-interactive zero-knowledge proofs for verifying sensor telemetry and ${\text{CO}}_2$ storage claims can be considered as an open research problem. These proofs constitute a crucial step in making the climate-positive blockchain feasible, secure and efficient.
    \item \textbf{Adversarial Scenarios:} We have described some adversarial situations and provided ways to mitigate them. Since the proposed approach involves connecting a physical process -- carbon capture and storage -- to a digital process -- a distributed consensus mechanism driving a blockchain -- novel adversarial scenarios arise, and need to be revealed and mitigated. For example, rogue DACCS facilities may fabricate additional fake identities and claim larger ${\text{CO}}_2$ storage amounts to increase their probability of winning the mining reward. Such behavior is difficult to defend against, but it can, in principle, be mitigated using a reputation system in which validator nodes can reduce the reputation score of a suspected rogue miner. For example, ${\text{CO}}_2$ storage claims from a fake location, or a location very close to an existing DACCS location, may raise suspicion of illicit activity. This approach is quite similar to reputation scoring used to prevent malicious behavior in some proof-of-stake mechanisms.
    \item \textbf{Novel Carbon Capture Mechanisms:} Though we have focused on DACCS as the carbon capture mechanism, the principles underlying the protocols apply to any alternative mechanism. It is an interesting research question to determine whether several carbon capture mechanisms can interoperate correctly, securely and fairly in the distributed consensus mechanism, and in its associated applications, such as driving cryptocurrency transactions.
   
   
\end{enumerate}

\subsubsection{\textbf{Governance and Oversight Considerations}}
\begin{enumerate}
\renewcommand{\labelenumi}{\alph{enumi}.}
    \item As we have described, the fact that we are interfacing a physical process (${\text{CO}}_2$ capture and storage) to a digital mechanism, creates new challenges of oversight that did not exist in previous (entirely digital) cryptocurrencies. How this oversight can be performed while preserving the benefits of decentralization and censorship resistance is a key challenge that needs to be addressed. 
    \item How can a governing body ensure that before permitting entry into the framework, the users are screened and determined to be interested investors who align with the larger ‘carbon negative’ mindset?
    \item How do you ensure that geographic migration to financially more lucrative operating regions does not make this a carbon positive endeavor?
    \item How can we ensure that there is a standard measure of the life cycle analysis (LCA) of any new added climate positive action, to be able to compare across processes?\\

\end{enumerate}